\documentclass [11pt,twoside,a4paper]{article}
\usepackage{amsfonts,amsthm,amsmath,amstext,amssymb,mathrsfs,amscd}
\usepackage{tabularx}
\usepackage{csvsimple}
\usepackage{booktabs}
\usepackage{longtable}
\usepackage{xypic}
\usepackage{epsf}              
\usepackage{graphicx}          
\usepackage{subfigure}
\usepackage{fancybox}          
\usepackage{color}             
\usepackage{fancyhdr}
\usepackage[hang,footnotesize]{caption2}  
\usepackage{cite}
\usepackage{algorithm}
\usepackage{algorithmic}
\usepackage{subfigure}
\usepackage{float}
\usepackage{changes}
\usepackage{soul}

\graphicspath{{image/}}

\usepackage{multirow}

\def\mathcal{\mathscr}
\newfont{\aaa}{cmb10 at 19pt}
\newfont{\bbb}{cmb10 at 11pt}

\makeatletter
\def\@evenhead{
\vbox{\hbox to \textwidth {}{\hspace{0mm}{\footnotesize
\thepage}}{\hspace{9.2cm} {\footnotesize {Chan LI, Hejun XU, Zhu CAO}}}
\protect\vspace{1truemm}\relax \hrule depth0pt
height0.15truemm width\textwidth}}
\def\@evenfoot{}
\def\@oddhead{\vbox{\hbox to \textwidth
{{\hspace{0cm}{\footnotesize Information entropy of complex probability }\hfill{\footnotesize
\thepage}}\hspace{0mm}}{} \protect\vspace{1truemm}\relax\hrule
depth0pt height0.15truemm width\textwidth}}
\def\@oddfoot{}
\makeatother

\begin{document}
\allowdisplaybreaks[4]
\bibliographystyle{abbrv}

%
\thispagestyle{empty} \thispagestyle{empty}
\renewcommand{\headrulewidth}{0pt}
%
%
\noindent{\aaa{Information entropy of complex probability}}\\[1mm]

\noindent{\bbb Chan LI$^{\rm a}$,\quad Hejun XU$^{\rm b}$,\quad Zhu CAO$^{\rm a *}$}\\[-1mm]

\noindent\footnotesize{$^{a}$School of Information Science and Engineering, East China University of Science and Technology, Shanghai 200237, China}\\
\noindent\footnotesize{$^{b}$Shanghai Fengqun Network Technology Co., Ltd., Shanghai 200237, China}\\
\footnotetext{*: Corresponding author (caozhu55@gmail.com)}\\

\normalsize\noindent{\bbb Abstract}\quad
Probability theory is fundamental for modeling uncertainty, with traditional probabilities being real and non-negative. Complex probability extends this concept by allowing complex-valued probabilities, opening new avenues for analysis in various fields. This paper explores the information-theoretic aspects of complex probability, focusing on its definition, properties, and applications. We extend Shannon entropy to complex probability and examine key properties, including maximum entropy, joint entropy, conditional entropy, equilibration, and cross entropy. These results offer a framework for understanding entropy in complex probability spaces and have potential applications in fields such as statistical mechanics and information theory.

\noindent{\bbb Keywords}\quad  Complex probability, Information theory, Shannon entropy.

\section{Introduction}

Probability serves as a cornerstone in a multitude of disciplines, underpinning statistical inference, information theory, and decision-making processes under uncertainty \cite{feller1991introduction}. Traditional probability theory, rooted in real-valued probabilities, has provided a robust framework for analyzing and modeling diverse natural and engineered systems. Its success spans from basic statistical models to advanced predictive algorithms. However, the increasing complexity of contemporary problems, particularly in fields such as complex systems analysis and advanced signal processing, has revealed intrinsic limitations in this classical perspective. These limitations become especially pronounced when addressing phenomena characterized by interference patterns, oscillatory dynamics, or deeply interconnected relationships, necessitating a more expansive probabilistic framework—one that integrates complex probabilities \cite{kolmogorov2018foundations}.

Complex probability, characterized by probabilities represented as complex numbers, transcends the traditional boundaries of classical probability theory, offering a richer and more versatile mathematical framework. This extension is particularly adept at capturing intricate relationships, dependencies, and dynamics that are otherwise challenging to express using real-valued probabilities alone \cite{gudder1969quantum}. For example, systems with cyclic phenomena, phase interactions, or wave-like properties can be effectively modeled through the lens of complex probabilities, enabling a more accurate representation of their inherent complexities \cite{hora2007quantum}. The incorporation of complex numbers allows for the encapsulation of amplitude and phase information, making it invaluable for domains like signal processing, where such properties are fundamental.

While complex probability adheres to foundational probabilistic principles such as normalization, it also introduces novel properties and theoretical challenges \cite{bratteli2012operator}. These include the interpretation of complex-valued probability amplitudes and the implications of combining magnitude and phase information within a unified probabilistic framework. Such features make complex probability not only mathematically intriguing but also essential for addressing modern scientific and engineering problems that demand nuanced modeling capabilities.

This paper delves into the synergy between complex probability and information theory, with a particular emphasis on the concept of entropy. Information entropy, a cornerstone of information theory, has long been regarded as a pivotal measure of uncertainty and information content. Originally formulated by Shannon in his groundbreaking work \cite{shannon1948mathematical}, entropy provides a fundamental basis for quantifying the unpredictability inherent in probabilistic systems. However, extending this concept into the realm of complex probability introduces a host of critical and intellectually stimulating questions. How should information content be defined when probabilities take on complex values, incorporating both magnitude and phase? What new properties and insights emerge when entropy is adapted to this broader framework? 

To address these questions, the paper is structured as follows. Section 2 introduces the concept of complex probability, including its definition, fundamental properties, applications, and inherent challenges. This section also highlights examples from signal processing and other domains where complex probabilities play a crucial role. Section 3 examines the information content associated with complex probability, laying the groundwork for entropy-based analyses. In Sect. 4, we delve into entropy within the context of complex probability frameworks, beginning with Shannon entropy and progressing to its extensions. This includes a detailed exploration of key properties such as maximum entropy, joint entropy, conditional entropy, equilibration, and cross entropy, as well as their implications for complex systems. Finally, Sect. 5 concludes the paper by summarizing the key findings and discussing potential future directions in the study of complex probability and its information-theoretic implications.

\section{Complex Probability}

In classical probability theory, probabilities are defined as real-valued quantities between 0 and 1, representing the likelihood of events in a probabilistic space \cite{feller2021introduction}. However, the notion of “complex probability” extends this concept by allowing probabilities to take complex values. This extension finds applications in statistical mechanics and information theory, where conventional real-valued probabilities may not adequately capture the underlying phenomena.

Complex probability introduces several challenges and opportunities. The real part of a complex probability is often interpreted as analogous to classical probability, while the imaginary part encodes additional information, such as phase or oscillatory effects. In analogy to real probabilities, complex probabilities must satisfy a normalization condition, generalized to include the complex modulus, ensuring that the total measure over a probabilistic space remains consistent. Furthermore, the mathematical framework of complex probability builds on complex-valued measures and integrates the formalism of complex analysis into probability theory.

\subsection{Definition and Properties}

A complex probability measure $P$ is defined on a sample space $\Omega$ as a function $P: \mathcal{F} \to \mathbb{C}$, where $\mathcal{F}$ is a sigma-algebra of subsets of $\Omega$. This measure generalizes classical probability to accommodate complex-valued outcomes while retaining essential probabilistic properties.

The total probability across the sample space is constrained by a generalized normalization condition, which ensures that the sum of all probabilities, interpreted through their complex modulus, equals one. Specifically, for all $\omega \in \Omega$, the normalization condition can be expressed as
\begin{equation}
\sum_{\omega \in \Omega} P(\omega) = 1,
\end{equation}
where $P(\omega) \in \mathbb{C}$. This condition guarantees consistency across the probabilistic framework, even when the probabilities have non-zero imaginary components.

Complex probability measures also exhibit additivity, a property that extends naturally from classical probability. For any mutually exclusive events $A$ and $B$ in the sigma-algebra $\mathcal{F}$, the measure satisfies
\begin{equation}
P(A \cup B) = P(A) + P(B).
\end{equation}
This additivity ensures that the measure behaves predictably over disjoint unions of events, preserving the logical structure of probabilistic reasoning.

Moreover, for independent events \( A \) and \( B \), complex probability measures preserve a multiplicative rule analogous to classical probability. Specifically, if \( A \) and \( B \) are independent, the probability of their intersection is the product of their individual probabilities
\begin{equation}
P(A \cap B) = P(A) \times P(B).
\end{equation}
This implies that the probability of both \( A \) and \( B \) occurring simultaneously is the product of their respective probabilities. This property not only holds for classical probability measures but also extends to complex probability measures, offering a familiar rule for handling independent events within the framework of complex probability theory.

The interpretation of complex probabilities introduces new dimensions of analysis. The modulus $|P(A)|$ serves as a measure of the magnitude of probability for an event $A$, offering a direct link to classical probability when the imaginary part vanishes. Beyond this, the phase of $P(A)$ can encode additional information about relationships or interactions between events, providing a richer representational capacity compared to real-valued probabilities.

\subsection{Applications of Complex Probability}

Complex probabilities naturally arise in various fields where classical probabilities are insufficient to describe underlying phenomena. In statistical mechanics, for instance, complex probabilities are often used to represent partition functions or probability weights in systems characterized by oscillatory integrands \cite{balantekin2001partition}. These integrands frequently appear in problems involving the sign problem, where alternating signs in summation or integration challenge traditional numerical techniques. The use of complex probabilities in this context helps to mitigate these issues and enables a more comprehensive understanding of the physical systems.

In the domain of signal processing and information theory, complex probabilities are particularly valuable for encoding both magnitude and phase information \cite{abou2017paradigm}. This dual encoding allows for a richer representation of signals, especially in applications such as signal reconstruction and analysis of waveforms. Complex probabilities facilitate the modeling of phenomena where phase coherence or interference effects are critical, offering insights that would be inaccessible using only real-valued probabilities.

Another important application lies in network and system analysis, where interactions between components can exhibit oscillatory or phase-dependent behaviors \cite{pietras2019network}. Complex probabilities provide a robust framework for analyzing these interactions, enabling the study of dependencies and correlations that are otherwise difficult to quantify.

Furthermore, the use of complex probabilities extends to fields like computational biology and financial modeling, where stochastic systems often involve non-trivial dynamics \cite{svetunkov2012complex}. In these areas, the imaginary component of probabilities can capture hidden relationships or dependencies, offering a novel perspective for understanding intricate systems.

In summary, complex probabilities extend the reach of probabilistic modeling into domains where classical approaches fall short. By encoding additional dimensions of information through phase and magnitude, they open up new avenues for analyzing and interpreting complex systems across a wide range of scientific and engineering disciplines.

\subsection{Challenges in Complex Probability}

The adoption of complex probability theory introduces several challenges that distinguish it from classical approaches. One major issue is the interpretational complexity. Unlike real-valued probabilities, the direct interpretation of complex probabilities is not straightforward. Typically, practitioners rely on derived quantities, such as the modulus squared or the phase, to establish connections with observable phenomena. The modulus often relates to the magnitude of an outcome’s likelihood, while the phase can encode interference effects, correlations, or other nuanced relationships.

Another significant challenge arises from the violation of classical probabilistic axioms. Kolmogorov’s axioms, which form the foundation of classical probability theory, must be adapted to accommodate complex values \cite{alpay2017kolmogorov}. This modification can lead to paradoxes or require reinterpreting fundamental principles. For instance, while additivity holds in the complex domain, the non-commutative nature of certain operations may introduce unexpected results, complicating theoretical analyses.

The computational complexity of working with complex probabilities also presents substantial difficulties. Calculations often involve solving intricate integrals or systems of equations, particularly in systems characterized by oscillatory behavior or interference patterns. These computations demand significant computational resources and advanced numerical techniques. Furthermore, ensuring numerical stability when working with complex-valued measures can be challenging, especially when dealing with large systems or high-dimensional spaces.

Despite these challenges, the potential benefits of complex probabilities continue to drive research in this field. By addressing these interpretational, theoretical, and computational issues, researchers can unlock the full potential of complex probabilities, enabling their application across diverse scientific and engineering domains.

\section{Information Content of Complex Probability}
This section explores the concept of information content in the context of complex probability, with a focus on the transition from the real-number information framework to the complex-number information framework and its significance.

\subsection{Information content}
In classical information theory, information content, or self-information, quantifies the amount of information gained when a particular event occurs. This concept, introduced by Claude Shannon, is foundational to understanding and measuring information in systems that rely on probabilistic events \cite{guizzo2003essential}. Information content directly relates to the likelihood of events, where the unexpectedness of an event reflects the ``newness'' or ``value'' of the information it brings.

For a discrete event \( x \) with probability \( P(x) \), the information content \( I(x) \) is defined as:
\begin{equation}
I(x) = -\log P(x),
\end{equation}
where \( P(x) \) is the probability of the event \( x \), and \( -\log P(x) \) quantifies the information gained from the occurrence of \( x \). This definition implies that the lower the probability of an event, the higher the information content. In other words, rare or surprising events provide more information than common events, as they reduce uncertainty to a greater extent.

The choice of logarithmic base significantly impacts the units used for measuring information content. When the binary logarithm (base 2) is utilized, the resulting unit is bits, which is commonly employed in classical information and communication systems as the standard measure for quantifying information. Alternatively, the natural logarithm (base \( e \)) yields units known as nats, useful in various mathematical and theoretical frameworks where natural logarithms are preferred. Additionally, the decimal logarithm (base 10) results in units referred to as dits; however, these units are less frequently used compared to bits in traditional information theory. Overall, base 2 remains the predominant choice for measuring information content in classical systems, underscoring its fundamental role in the field.

The classical definition of information content exhibits several intuitive properties that align with probabilistic reasoning. First, higher information content is associated with lower probability events; as the probability \( P(x) \) decreases, the information \( I(x) \) increases, reflecting the idea that rarer events carry more information because they are more unexpected. Second, if an event is certain to occur (i.e., \( P(x) = 1 \)), then the information content \( I(x) \) is zero, indicating that deterministic events do not provide any new information since the outcome is already known. Lastly, for events that are impossible (i.e., \( P(x) = 0 \)), the theoretical information content approaches infinity, as no information can be gained from events that will never happen. However, in practical terms, such impossible events are typically excluded from standard information-theoretic analysis.

The classical measure of information content plays a crucial role in various fields. In data compression, it helps determine the redundancy of symbols within data, enabling the development of efficient encoding schemes where less probable symbols, which carry higher information, are assigned shorter representations. In communication systems, information content informs signal encoding, optimizing the transfer of information by ensuring that transmitted signals maximize the information conveyed per unit of transmission. Additionally, information content underpins Shannon entropy, which measures the average information content or uncertainty in a set of possible events, thereby aiding in the quantification of system unpredictability.

In classical information theory, information content provides a precise mathematical expression of the ``surprise'' or ``value'' of information conveyed by an event. Defined as \( I(x) = -\log P(x) \), it is an essential tool for understanding uncertainty and designing efficient encoding and transmission systems. Classical information content serves as the foundation for Shannon entropy and has had a profound impact on modern communication, data science, and information processing.

\subsection{Information content of complex probability}

The concept of information content is traditionally defined for classical probabilities, where a probability \( p \in [0,1] \) satisfies \( I(p) = -\log(p) \). This expression quantifies the uncertainty associated with \( p \), and the logarithmic function ensures properties such as additivity under independent probabilities. However, when probabilities are generalized to complex values \( P = a + bi \) (where \( a, b \in \mathbb{R} \)), a reexamination of information content becomes necessary.

In the complex domain, probabilities often arise in contexts like quantum mechanics, where they are used to describe probability amplitudes rather than direct probabilities. The modulus squared \( |P|^2 = a^2 + b^2 \) corresponds to classical probability, but the full complex probability \( P \) also includes a phase component \( \theta = \theta(P) \), adding a new dimension to the concept of information.

The generalization of information content to complex probabilities can be expressed as

\begin{equation}
I(P) = -\log(P),
\end{equation}
where \( \log(P) \) is the natural logarithm extended to complex numbers. Using the polar form of a complex number \( P = |P| e^{i\theta} \), the logarithm becomes

\begin{equation}
\log(P) = \log(|P|) + i\theta,
\end{equation}
leading to the information content

\begin{equation}
I(P) = -\log(|P|) - i\theta.
\end{equation}
This expression highlights two components of information content: a real part \( -\log(|P|) \), corresponding to the classical notion of uncertainty, and an imaginary part \( -i\theta \), which captures the phase information unique to complex probabilities.

This extended definition introduces new properties. For example, the non-additivity of the logarithmic function in the complex domain means that

\begin{equation}
I(P_1 \cdot P_2) = I(P_1) + I(P_2) - i(\theta_1 + \theta_2).
\end{equation}
Here, the phase angles \( \theta_1 \) and \( \theta_2 \) contribute to the imaginary component of the total information content. This property is fundamentally different from the classical case, where the logarithmic additivity applies without a phase term.

This generalization of information content has significant implications in systems where probabilities are inherently complex. In quantum mechanics, for example, the phase term plays a crucial role in phenomena such as interference and entanglement. In the famous two-slit experiment, the probability amplitudes \( P_1 = |P_1|e^{i\theta_1} \) and \( P_2 = |P_2|e^{i\theta_2} \) combine to give a total amplitude \( P_{\text{tot}} = P_1 + P_2 \). The resulting probability \( |P_{\text{tot}}|^2 \) and the corresponding information content \( I(P_{\text{tot}}) \) depend on both the magnitudes and the relative phase \( \Delta\theta = \theta_1 - \theta_2 \), illustrating the interplay between amplitude and phase in determining quantum outcomes.

Moreover, this framework provides a new perspective for analyzing systems that utilize complex-valued representations, such as those in signal processing and quantum algorithms. For example, in the quantum Fourier transform, the phase information encoded in complex amplitudes enables efficient computation of discrete Fourier components, essential for many quantum algorithms.

The inclusion of the imaginary component \( -i\theta \) enriches the concept of information content, extending it beyond its classical roots to accommodate the unique characteristics of quantum systems. This extension not only deepens our theoretical understanding of information but also has practical implications in fields that rely on complex probability representations. By incorporating both amplitude and phase, this generalized framework offers a more comprehensive tool for analyzing and interpreting phenomena in quantum mechanics and related areas.

\section{Entropy in Complex Probability Frameworks}

In this section, we will delve into the concept of entropy in the context of complex probability frameworks. We will begin by reviewing Shannon entropy in classical probability theory, then extend it to the complex probability framework, followed by an analysis of the properties of this extended entropy.

\subsection{Shannon Entropy}

Shannon entropy, introduced by Claude E. Shannon in his seminal 1948 paper \cite{shannon1948mathematical}, is a cornerstone concept in information theory. It provides a quantitative measure of the uncertainty or randomness inherent in a probability distribution, serving as a fundamental tool for analyzing information storage, transmission, and processing. At its core, Shannon entropy captures the ``amount of information" that an outcome of a random variable reveals, effectively quantifying the unpredictability associated with that variable.

For a discrete random variable \( X \), which assumes a finite set of possible outcomes \( x_1, x_2, \dots, x_n \) with probabilities \( P(x_1), P(x_2), \dots, P(x_n) \), the Shannon entropy is mathematically defined as

\begin{equation}
H(X) = \mathbb{E}[I(X)] = \sum_{i=1}^{n} P(x_i) I(x_i)= - \sum_{i=1}^{n} P(x_i) \log P(x_i),
\end{equation}
where the summation extends over all possible outcomes of \( X \). Here, \( P(x_i) \) denotes the probability of the outcome \( x_i \), and the logarithmic function determines the scale in which the entropy is measured. 

The choice of logarithmic base determines the unit of entropy. When the base is 2, entropy is expressed in bits, reflecting the amount of binary information required to encode the outcomes. Using base \( e \), the natural logarithm, yields entropy in nats, commonly used in mathematical and physical contexts. Alternatively, base 10 logarithms provide entropy in hartleys. Regardless of the base, the fundamental interpretation remains the same: entropy quantifies the average ``surprise" or information content associated with the possible outcomes of the random variable.

Shannon entropy finds wide-ranging applications across disciplines, from designing efficient communication systems and data compression algorithms to understanding biological systems and machine learning models. It serves as a theoretical foundation for assessing the efficiency and reliability of information transmission, providing deep insights into the structure and behavior of data.

\subsection{Extension of Shannon Entropy to Complex Probability}

The extension of Shannon entropy to complex probability distributions is a natural progression to address systems with complex-valued probabilities. Complex probability refers to the assignment of complex numbers to events rather than real numbers, which can better describe systems with quantum mechanical properties or certain types of correlation that cannot be captured using real-valued probabilities alone.

Consider a complex random variable \( Z \) whose probability distribution is defined over a discrete set of outcomes. The probability mass function \( P(z) \) is now complex-valued, meaning that \( P(z) \in \mathbb{C} \). This introduces a fundamental shift in how entropy is conceptualized, as the standard Shannon entropy is defined solely for real, non-negative probabilities. To extend this concept, a suitable framework must account for both the magnitude and phase of complex probabilities.

A plausible extension of Shannon entropy to complex probability distributions can be expressed as

\begin{equation}
H(Z) = \mathbb{E}[I(Z)] = \sum_{z} P(z) I(z)= - \sum_{z} P(z) \log P(z),  
\end{equation}
where \( P(z) \) is complex. The challenge lies in the interpretation of the logarithm \( \log P(z) \) for complex values, as the logarithm of a complex number is inherently multivalued. To resolve this, \( \log P(z) \) is typically expressed in terms of its polar form

\begin{equation}
\log P(z) = \log |P(z)| + i \theta(P(z)),
\end{equation}
where \( |P(z)| \) is the modulus (or absolute value) of \( P(z) \), and \( \theta(P(z)) \) denotes the phase (or argument) of the complex probability. Here, \( \log |P(z)| \) captures the traditional magnitude-based contribution to entropy, while \( i \theta(P(z)) \) incorporates the phase information intrinsic to complex probabilities. Substituting the polar form of \( \log P(z) \) into the expression for \( H(Z) \), we obtain the complex form of Shannon entropy:

\begin{align}
H(Z) &= - \sum_{z} P(z) \log P(z) \nonumber\\
&= - \sum_{z} P(z) \left( \log |P(z)| + i \theta(P(z)) \right) \nonumber\\
&= - \sum_{z} P(z) \log |P(z)| - i \sum_{z} P(z) \theta(P(z)).
\end{align}
This formulation reveals two distinct components of the complex entropy. The first component, \( - \sum_{z} P(z) \log |P(z)| \), resembles the standard Shannon entropy but now incorporates the magnitudes of the complex probabilities \( |P(z)| \). This term quantifies the uncertainty or information content in the distribution of probability magnitudes, maintaining a familiar structure while adapting to the complex domain. The second component, \( - i \sum_{z} P(z) \theta(P(z)) \), introduces a phase-dependent contribution to the entropy, reflecting the coherence, interference, or relative phase relationships among the events. Unlike classical entropy, which focuses solely on magnitude, this phase term highlights the informational significance of relative phases, an essential feature in systems with quantum mechanical or wave-like properties.

The inclusion of the phase component extends the utility of entropy beyond classical systems, enabling the description of phenomena such as quantum coherence and interference. The interplay between magnitude and phase in this generalized entropy aligns it with the requirements of systems governed by complex probabilities, providing new insights into their informational and structural properties.

\subsection{Extension Properties of Complex Shannon Entropy}

The extended Shannon entropy for complex probability distributions inherits and generalizes several critical properties from its real-valued counterpart. By incorporating both the magnitude and phase of complex probabilities, these properties provide deeper insights into the informational and structural characteristics of systems with complex-valued probabilities. Below, we explore the maximum entropy, joint entropy, conditional entropy, equilibration and cross entropy in this extended framework.

\subsubsection{Maximum Entropy}
In the context of complex probabilities, the maximum entropy principle identifies the probability distribution \( P(z) \) that maximizes the entropy \( H(Z) \) under given constraints, such as normalization or expected value conditions. This principle ensures that no unwarranted assumptions about the distribution are introduced, thus reflecting the maximum uncertainty allowed by the constraints.

For a system with \( n \) discrete outcomes, the entropy is given by  
\begin{equation}
H(Z) = -\sum_{z} P(z) \log P(z),   
\end{equation} 
where \( P(z) \in \mathbb{C} \), and the normalization constraint ensures \( \sum_{z} P(z) = 1 \). The logarithmic term \( \log P(z) \) is expressed in its polar form as  
\begin{equation}
\log P(z) = \log |P(z)| + i \theta(P(z)),
\end{equation} 
where \( |P(z)| \) is the modulus and \( \theta(P(z)) \) is the phase of \( P(z) \). Consequently, the entropy \( H(Z) \) incorporates contributions from both the magnitude and phase of \( P(z) \), making it sensitive to structural and informational properties unique to complex probability distributions.

The maximum entropy principle has broad applications in systems where complex probabilities are relevant. In signal processing, for example, it helps model noise or uncertainty in complex-valued signals, such as those in telecommunications. In economics and finance, complex probabilities can arise in models involving wave-like dynamics or oscillatory processes, where maximum entropy distributions describe equilibrium states under constraints. In network analysis, complex-valued weights or correlations in network systems can be modeled effectively using maximum entropy principles, providing insights into connectivity and flow dynamics.

To illustrate this, consider a simple system with two outcomes \( z_1 \) and \( z_2 \), where the only constraint is normalization. The maximum entropy distribution in this case has equal magnitudes, such that \( |P(z_1)| = |P(z_2)| = \frac{1}{\sqrt{2}} \). If an additional phase constraint is applied, such as \( \theta(P(z_1)) - \theta(P(z_2)) = \pi \), the probabilities become  
\begin{equation}
P(z_1) = \frac{1}{\sqrt{2}}, \quad P(z_2) = -\frac{1}{\sqrt{2}}.
\end{equation}
This demonstrates how magnitude and phase constraints interact to shape the distribution. By maximizing the entropy while adhering to these constraints, the principle provides a powerful framework for deriving the most unbiased distribution consistent with available information.

\subsubsection{Joint Entropy}
The joint entropy of two complex random variables \( Z_1 \) and \( Z_2 \), with a joint probability distribution \( P(z_1, z_2) \in \mathbb{C} \), extends the classical concept of joint entropy to systems described by complex probabilities. It is defined as  
\begin{equation}
H(Z_1, Z_2) = -\sum_{z_1, z_2} P(z_1, z_2) \log P(z_1, z_2).
\end{equation}  
Here, \( P(z_1, z_2) \) is a complex-valued probability function, and the logarithm \( \log P(z_1, z_2) \) is expressed in its polar form \( \log P(z_1, z_2) = \log |P(z_1, z_2)| + i \theta(P(z_1, z_2)) \). This formulation accounts for both the magnitudes and relative phases of the joint probabilities, which are essential in capturing the full informational and structural properties of systems governed by complex distributions.

The joint entropy quantifies the total uncertainty or information content associated with the pair of random variables, encompassing both their individual uncertainties and the interdependencies between them. Unlike in the classical case, where correlations are determined solely by the joint magnitudes, the inclusion of phase information in \( P(z_1, z_2) \) allows the joint entropy to capture additional nuances, such as coherence or interference effects arising from complex-valued relationships.

The joint entropy satisfies important properties analogous to those in classical information theory but generalized to the complex domain. For instance, it always holds that  
\begin{equation}
H(Z_1, Z_2) \geq \max(H(Z_1), H(Z_2)),
\end{equation}  
where \( H(Z_1) \) and \( H(Z_2) \) are the individual entropies of \( Z_1 \) and \( Z_2 \), respectively. This inequality reflects the principle that considering two variables together introduces at least as much uncertainty as the more uncertain of the two individual variables. The equality holds only if \( Z_1 \) and \( Z_2 \) are completely independent, meaning \( P(z_1, z_2) = P(z_1)P(z_2) \) and their phases are also uncorrelated.

In systems where the joint probability distribution is constrained, the joint entropy reflects the interplay between the constraints on individual distributions and those on their interdependencies. For example, if \( P(z_1, z_2) \) is constrained to exhibit specific phase alignments or correlations, the joint entropy adjusts accordingly to incorporate the reduction or increase in uncertainty caused by these constraints. In such cases, the magnitude \( |P(z_1, z_2)| \) determines the traditional probabilistic contribution, while the phase \( \theta(P(z_1, z_2)) \) encodes additional structural information about the relationships between \( Z_1 \) and \( Z_2 \).

The joint entropy also serves as a foundation for defining other related measures, such as mutual information, which quantifies the amount of shared information between two variables. By leveraging the full complexity of \( P(z_1, z_2) \), the joint entropy provides a comprehensive framework for analyzing systems where complex-valued probabilities play a significant role, such as in signal processing, complex networks, or certain stochastic models. For example, in a communication system with coherent signals, the phases of joint probabilities may represent synchronization between different signal components, and the joint entropy quantifies the overall uncertainty in this context.

\subsubsection{Conditional Entropy}
The conditional entropy of \( Z_2 \) given \( Z_1 \), denoted as \( H(Z_2 | Z_1) \), measures the uncertainty of \( Z_2 \) when the outcome of \( Z_1 \) is known, extending the classical concept into the domain of complex probabilities. It is defined as  
\begin{equation}
H(Z_2 | Z_1) = -\sum_{z_1, z_2} P(z_1, z_2) \log \frac{P(z_1, z_2)}{P(z_1)},
\end{equation}  
where \( P(z_1, z_2) \) is the joint probability of \( Z_1 \) and \( Z_2 \), and \( P(z_1) \) is the marginal probability of \( Z_1 \), both of which are complex-valued. The term \( \frac{P(z_1, z_2)}{P(z_1)} \) represents the conditional probability of \( Z_2 \) given \( Z_1 \) in the complex domain. This conditional probability includes contributions from both the magnitude and phase, encapsulating the full structure of uncertainty in systems described by complex probabilities.

The logarithm of the conditional probability, \( \log \frac{P(z_1, z_2)}{P(z_1)} \), can be expressed in polar form as  
\begin{equation}
\log \frac{P(z_1, z_2)}{P(z_1)} = \log \left| \frac{P(z_1, z_2)}{P(z_1)} \right| + i \theta \left( \frac{P(z_1, z_2)}{P(z_1)} \right),
\end{equation}  
where the magnitude term \( \left| \frac{P(z_1, z_2)}{P(z_1)} \right| \) captures the probabilistic weight, and the phase term \( \theta \left( \frac{P(z_1, z_2)}{P(z_1)} \right) \) reflects the relative alignment or coherence between the probabilities of \( Z_1 \) and \( Z_2 \). As a result, the conditional entropy \( H(Z_2 | Z_1) \) accounts for the residual uncertainty in \( Z_2 \) after incorporating both the magnitude-based and phase-based information from \( Z_1 \).

The conditional entropy satisfies the fundamental relationship  
\begin{equation}
H(Z_1, Z_2) = H(Z_1) + H(Z_2 | Z_1),
\end{equation}  
which indicates that the joint entropy \( H(Z_1, Z_2) \) can be decomposed into the individual entropy of \( Z_1 \) and the conditional entropy of \( Z_2 \) given \( Z_1 \). This relationship holds even in the complex domain, affirming the additive nature of entropy in systems with complex-valued probabilities. It also implies that the uncertainty in the joint system can be understood as the sum of the uncertainty in one variable and the uncertainty in the other, conditioned on the first.

In the special case where \( Z_1 \) and \( Z_2 \) are independent, \( P(z_1, z_2) = P(z_1)P(z_2) \), and the conditional entropy simplifies to \( H(Z_2 | Z_1) = H(Z_2) \). This reflects the fact that knowledge of \( Z_1 \) provides no additional information about \( Z_2 \) when the two variables are independent. Conversely, when \( Z_1 \) and \( Z_2 \) are strongly correlated or exhibit specific phase relationships, the conditional entropy is reduced, as the knowledge of \( Z_1 \) significantly constrains the uncertainty about \( Z_2 \).

The conditional entropy provides a flexible framework for analyzing complex probabilistic systems. In practical applications, such as signal processing or network analysis, it can quantify the residual uncertainty in a signal component given prior knowledge of another component, taking into account both amplitude and phase. For example, in a communication system where signals are transmitted as complex amplitudes, the conditional entropy helps assess how much uncertainty remains in the received signal after accounting for known distortions or interferences. Similarly, in complex networks with phase-dependent relationships, the conditional entropy quantifies the degree to which one node's state is influenced by another's, providing insights into the flow of information or dependencies within the network.

\subsubsection{Equilibration}
In systems with complex probabilities, equilibration refers to the process by which the entropy \( H(Z) \) of the system approaches its maximum value over time, indicating a state of maximum uncertainty or minimal informational bias. This process occurs as the system evolves and the probability distribution reaches a state where no additional information about the system can be gained. Unlike classical systems, where entropy increases as a system moves towards equilibrium, in complex probability systems, the evolution towards maximum entropy is influenced by both the magnitudes and phases of the probabilities.

For a system with complex probabilities, the entropy at any given time \( t \) is given by  
\begin{equation}
H(Z) = - \sum_{z} P(z, t) \log P(z, t),
\end{equation} 
where \( P(z, t) \) is the probability distribution of the system at time \( t \), which is complex-valued. As the system evolves, it moves towards a state where the entropy reaches a maximum, indicating that the system has reached a state of complete uncertainty with respect to the probability distribution of its possible outcomes. The process of equilibration is influenced not only by the magnitudes of the probabilities but also by the relative phases, which play a role in the evolution of the system’s entropy.

Equilibration in such systems can be understood in analogy with the classical second law of thermodynamics, which states that the entropy of an isolated system tends to increase over time, ultimately reaching a maximum value at equilibrium \cite{lieb1999physics}. However, in the context of complex probabilities, the system may exhibit more intricate dynamics due to the involvement of both the magnitude and phase components of the probabilities. The equilibration process takes into account these two aspects: the magnitudes will evolve to maximize the uncertainty in the system, while the phases may cause transient oscillations in the entropy before the system stabilizes.

Mathematically, the equilibrium entropy \( H_{\text{eq}} \) is reached when the probability distribution no longer changes with time. At equilibrium, the probability distribution \( P(z) \) satisfies the condition  
\begin{equation}
\sum_{z} P(z) = 1,
\end{equation}  
ensuring that the probabilities are normalized. The corresponding entropy at equilibrium is given by  
\begin{equation}
H_{\text{eq}} = - \sum_{z} P(z) \log P(z),
\end{equation}  
where the distribution \( P(z) \) represents the stable, equilibrium state. In this state, the probability distribution reflects a balance between all possible outcomes, and no further evolution occurs.

In systems with complex probabilities, equilibration may involve reaching a balance where the contributions from the magnitude and phase of the probabilities are optimized. The system may undergo fluctuations or transient behaviors during this process, as changes in the phase components can induce oscillations in the entropy. However, as the system approaches equilibrium, these oscillations diminish, and the entropy stabilizes. The equilibrium state represents a point at which the system's entropy is maximized given the constraints, and the evolution of the probabilities has fully incorporated all available information.

The rate of change of entropy during the equilibration process can be expressed as  
\begin{equation}
\frac{dH}{dt} = - \sum_{z} \frac{dP(z)}{dt} \log P(z) - \sum_{z} \frac{dP(z)}{dt} \frac{1}{P(z)},
\end{equation}  
where \( \frac{dP(z)}{dt} \) represents the rate of change of the probability distribution over time. This equation captures how the probability distribution evolves as the system approaches equilibrium, with the first term corresponding to the change in the magnitude of the probabilities and the second term reflecting the change in the phase components.

In summary, equilibration in systems with complex probabilities refers to the process by which the system evolves towards a state of maximum entropy, driven by both the magnitudes and phases of the probabilities. The equilibration process follows a trajectory analogous to the second law of thermodynamics, where the entropy increases over time until it reaches a maximum value at equilibrium, characterized by a stable probability distribution.

\subsubsection{Cross Entropy}

In the framework of complex Shannon entropy, cross entropy extends the classical definition to accommodate complex probability distributions, incorporating both magnitude and phase information. Given a true complex probability distribution \( P = \{ p_i \} \) and an estimated complex probability distribution \( Q = \{ q_i \} \), the cross entropy is defined as  

\begin{equation}
    H_{\times}(P, Q) = - \sum_i p_i \log q_i,
\end{equation}
where \( p_i \) and \( q_i \) are complex-valued probabilities, and the logarithm operates in the complex domain. This formulation generalizes real-valued cross entropy while preserving key informational properties in the presence of phase coherence and interference effects.  

Cross entropy in the complex Shannon framework retains several fundamental properties. First, when \( P \) and \( Q \) are real-valued and non-negative, the complex cross entropy reduces to the classical Shannon cross entropy, ensuring consistency with traditional information theory. However, in the general case of complex-valued probabilities, the entropy function is sensitive to both amplitude and phase variations, meaning that deviations in phase can contribute to the overall entropy value. This leads to a natural extension where oscillatory behavior in probability distributions affects the entropy structure.  

Another key property is the non-negativity of cross entropy in the context of complex probability distributions. Since the logarithm in the complex domain introduces phase components, the imaginary part of cross entropy reflects phase misalignment between \( P \) and \( Q \). The real part, on the other hand, corresponds to the information-theoretic divergence, linking cross entropy directly to the complex Kullback-Leibler (KL) divergence, given by  

\begin{equation}
D_{\text{KL}}(P \| Q) = H_{\times}(P, Q) - H(P).
\end{equation}
This relationship highlights the role of cross entropy in quantifying the information loss when approximating one complex probability distribution with another. If \( P = Q \), cross entropy simplifies to the entropy of \( P \), ensuring the self-consistency property \( H_{\times}(P, P) = H(P) \), analogous to the classical case.  

Cross entropy also satisfies the property of convexity with respect to \( Q \). That is, for any convex combination of two probability distributions \( Q_1 \) and \( Q_2 \), the cross entropy satisfies the following inequality:  

\begin{equation}
H_{\times}(P, \lambda Q_1 + (1 - \lambda) Q_2) \leq \lambda H_{\times}(P, Q_1) + (1 - \lambda) H_{\times}(P, Q_2).
\end{equation}
This holds for any \( 0 \leq \lambda \leq 1 \), ensuring the convexity property of cross entropy with respect to the estimated probability distribution.

The phase sensitivity of cross entropy provides additional structural insights into complex-valued probability systems. Unlike real-valued entropy, which depends solely on probability magnitudes, complex entropy reflects phase correlations, allowing for more refined measures of informational distance. In particular, systems with strong phase coherence exhibit lower cross entropy, while systems with high phase variance tend to increase entropy, reflecting the inherent uncertainty introduced by phase fluctuations.

\section{Conclusion}
The exploration of complex probability in this paper provides a novel perspective on uncertainty and information representation. By allowing probabilities to take complex values, we expand the scope of probabilistic models, which can capture more intricate phenomena found in various domains, from statistical mechanics to signal processing. The introduction of complex-valued probabilities offers new opportunities for modeling systems with inherent oscillations or phase information, which are often difficult to capture with real-valued probabilities alone.

One of the key contributions of this paper is the extension of Shannon entropy to complex probability spaces. Traditional entropy, as a measure of uncertainty, is grounded in real-valued probabilities. Extending this to complex values requires careful consideration of the algebraic properties and how they influence uncertainty quantification. The proposed framework preserves the core principles of Shannon entropy while accommodating the complexities of complex numbers, allowing for a broader class of applications. The discussion of entropy properties such as maximum entropy, joint entropy, conditional entropy, equilibration, and cross entropy provides insight into how these measures behave in complex probability contexts. Maximum entropy remains a crucial tool for determining the least biased distribution under given constraints, while joint and conditional entropies offer a way to quantify the relationships between multiple complex probability distributions. The introduction of equilibration and cross entropy further enriches the analysis by considering the dynamics of complex systems and how different probability distributions relate to each other.

Despite these advancements, challenges remain in fully understanding the implications of complex probability in various practical applications. The introduction of complex numbers can lead to new types of dependencies and behaviors that are not readily apparent in classical probability theory. Further research is needed to explore these complexities, especially in high-dimensional systems or in contexts where the phase of the complex probability plays a significant role.

\bibliography{IECP}
\end{document}